%Paper: hep-th/9209073
%From: SALEUR%YALPH2.bitnet@yalevm.ycc.yale.edu
%Date: 21 Sep 1992 11:37:35 -0500 (EST)

\input harvmac.tex
\Title{\vbox{\baselineskip12pt\hbox{YCTP-P35-1992}
                \hbox{UTTG-21-92}}} {\vbox{\centerline{Reidemeister Torsion}
\vskip2pt\centerline{the Alexander Polynomial}
\vskip2pt\centerline{and $U(1,1)$ Chern Simons Theory}}}

\centerline{L.Rozansky$^*$ and H. Saleur$^\dagger$}
\bigskip\centerline{$^*$Department of Physics}
\centerline{University of Texas at Austin}
\centerline{Austin, TX 78712}
\bigskip\centerline{$^\dagger$Department of Physics}
\centerline{Yale University}
\centerline{New Haven, CT 06519}
\bigskip

\vskip .3in
Abstract: We show
that the $U(1,1)$ (super) Chern Simons theory
 is one loop exact. This provides a direct proof of the relation between
the Alexander polynomial and  analytic and Reidemeister torsion.
 We then proceed to compute
explicitely the torsions of
Lens spaces and Seifert manifolds using surgery and the $S$ and $T$ matrices of
the $U(1,1)$ Wess Zumino Witten model recently determined, with complete
agreement with known results.
$U(1,1)$ quantum field theories and the Alexander polynomial provide thus "toy"
models with a non trivial topological content,
where all ideas put forward by Witten for $SU(2)$ and the Jones
polynomial can be explicitely checked, at finite $k$. Some simple but
presumably generic aspects of non compact groups, like the modified relation
between Chern Simons and Wess Zumino Witten theories, are also illustrated.
We comment on the
closely related case of $GL(1,1)$.

\bigskip

\vfill\eject

Quantum field theory plays an increasing role in the study of topological
invariants in low dimensions \ref\Sch{A.Schwarz, Lett. Math. Phys. 2 (1978)
247.}. In his seminal paper \ref\WiI{E.Witten, Comm.
Math. Phys. 121 (1989) 351.} Witten has shown how $SU(2)$ Chern Simons theory
gave rise to the Jones polynomial for links in $S^3$, to new 3-manifold
invariants, and to new "Jones" invariants of links in manifolds. Subsequent
developments have taken place in several directions. In particular
3-manifold invariants have
been defined more rigorously
 \ref\RT{N.Yu.Reshetikin, V.Turaev, Invent. Math. 103 (1991)
547.},\ref\Cr{L.Crane, Comm. Math. Phys. 135 (1991) 615.},\ref\K{T.Kohno,
"Topological Invariants for Three Manifolds Using Representations of Mapping
class Groups", Nagoya Univ. Preprint 6 (1990).} and their
 interrelations studied \ref\Piu{S.Piunikhin, "Reshetikhin-Turaev and
Kontsevich-Kohno-Crane 3-Manifolds Invariants Coincide", to appear in J.Knot
Theory.} . Also
3-manifold invariants have been explicitely obtained via surgery and the $S$
and $T$ matrices of the Wess Zumino Witten theories, either numerically
 \ref\Fg{D.Freed, R.Gompf, Comm. Math. Phys. 141 (1991) 79.} or analytically
\ref\Lisa{L.Jeffrey, "On some aspects of Chern
Simons theory", Oxford Univ. Ph.D. Thesis, and
Comm. Math. Phys. 147 (1992) 563.}.
 Of particular interest in that case
has been the large $k$ expansion \WiI , \ref\AXS{S.Axelrod, I.M.Singer., "Chern
Simons Perturbation Theory", in Proceedings of XXth International Conference of
Differential Geometrical Methods in Theoretical Physics, S.Catto and
 A.Rocha Eds., World Scientific (1991).}
 in which  the
analytic torsion \ref\RayS{D.B.Ray, I.M.Singer, Adv. in Math. 7 (1971) 145.},
 equal to the Reidemeister torsion
\ref\Mueller{W.Mueller, Adv. in Math. 28 (1978) 233.}, \ref\Cheeger{J.Cheeger,
Proc. Nat. Acad. Sci. USA 74 (1977) 2651.}, appears. Of course the
complete 3-manifolds invariants in the $SU(2)$ case contain more topological
information than the torsion.

Following these recent works on the Jones polynomial,
 the older Alexander Conway  \foot{By Alexander polynomial we refer
to the original invariant defined up to powers of the variable $t$. By
Alexander Conway polynomial we refer to its normalized version. See
\ref\K{L.H.Kauffman, "Formal Knot Theory", Princeton University Press (1983).}
and references therein.} polynomial
 \ref\Rolf{D.Rolfsen, "Knot and Links", Publish or Perish Press (1976).},
\ref\Kau{L.Kauffman, "On Knots", Princeton University Press (1987).}
 has been reconsidered and partly
 put in this more modern perspective \ref\RsI{L.Rozansky, H.Saleur,
Nucl.  Phys. B376 (1992) 461}, \ref\RsII{L.Rozansky, H.Saleur,
"$S$ and $T$ Matrices for the Super $U(1,1)$ WZW Model", Preprint
YCTP-P10 (1992), to appear in Nucl. Phys. B.}.
  It appears interesting to do so in order, for
instance, to gain more insight on the topological meaning of the Jones
polynomial itself, or to have at hand a case where simple computations
can be made and the whole construction tested. In \RsII\ , the $S$ and
$T$ matrices of the $U(1,1)$ Wess Zumino Witten model have been
computed, and it has been shown how surgery allows to compute the
multivariable Alexander Conway polynomial of links in $S^3$. The purpose
 of this
letter is to study invariants of manifolds and their relations with
torsion.

Relations between torsion and the Alexander invariant are well known: in
 \ref\Turaev{V.Turaev, Russ. Math. Surv. 41
 (1986) 199.}
Turaev, following  previous work by Milnor
 \ref\Milnor{J.Milnor, Ann. of Math. 76
(1962), 137.}, has  established
the equivalence of the Alexander polynomial of links in 3-manifolds and the
Reidemeister torsion of the complements of those links as well as the torsion
of the manifolds themselves. Recalling the above mentioned equality of
Reidemeister and analytic torsion,
we  complete here
 the relations of these
three mathematical objects by showing that the  Chern Simons theory
based on $U(1,1)$ is one-loop exact, thus providing a direct relationship
 between
the Alexander polynomial and analytic torsion.

We then compute Alexander polynomials of links in a variety of Lens spaces and
Seifert manifolds using the $S$ and $T$
matrices of the $U(1,1)$ Wess Zumino Witten model, and extract from them
(at finite $k$) Reidemeister torsions.

\bigskip

In all this paper we use notations of \RsI\ .
Recall the expression of the "classical" $U(1,1)$ Chern Simons
 action in components
\eqn\action{S_{\hbox{cl}}={k\over
2\pi}\int\epsilon_{\mu\nu\rho}\left(A^{\mu}_N\partial^{\nu}
A^{\rho}_E
+A^{\mu}_{\psi}(\partial^{\nu}+A^{\nu}_E)A^{\rho}_{\psi^{+}}\right)d^3x}
where we have set
\eqn\connection{A^{\mu}=A^{\mu}_EN+A^{\mu}_NE+
A^{\mu}_{\psi}\psi^{+}-A^{\mu}_{\psi^{+}}
\psi}
Recall that the generators satisfy relations
\eqn\rel{\{\psi,\psi^+\}=E,\ [N,\psi^+]=\psi^+,\ [N,\psi]=-\psi,\
E\hbox{
is central}}
We consider first
 purely bosonic flat connections,
determined by
\eqn\flat{F^{\mu\nu}_E=
\partial^{\mu}A^{\nu}_E-\partial^{\nu}A^{\mu}_E=0,\
F^{\mu\nu}_N=\partial^{\mu}A^{\nu}_N-
\partial^{\nu}A^{\mu}_N=0}
They are labelled by maps from $H_1({\cal M})$ into $U(1)_N\times
U(1)_E$ ($H_1$ arises instead of $\Pi_1$, $U(1)$ being abelian). Since the
bosonic generators commute among themshelves, flat connections are always
reducible.

The one loop approximation to 3-manifold invariants
 is obtained as in the standard case. Call
$A^{(\alpha)}$ a complete set of gauge equivalent flat connections on the
manifold ${\cal M}$. Assume
 that there is a finite number of them, and for the moment that none
of them are reducible. Expanding the fields around $A^{(\alpha)}$ we
get a quadratic action, to be supplemented by proper gauge fixing
terms, after a metric is picked on ${\cal M}$.  The $A_E$ and $A_N$
 degrees of freedom commute with $A^{(\alpha)}$ and the resulting
integrations do not depend on it.  We get in the notations of
 \ref\WiI{E.Witten, Comm. Math. Phys. 121 (1989) 351.}
\eqn\firstpart{(\hbox{det}(\Delta))^2\over|\hbox{det}(L_-)|}
(each
acting on "one colored" objects, as opposed to the three colors for
$SU(2)$). For $A_{\psi}$ and $A_{\psi^{+}}$ we get, due to statistics,
the inverse of the above result, but without absolute value and with
the operators $\Delta$ and $L_-$ being twisted by the $A_E^{(\alpha)}$
 part of the flat connection (the $A_N^{(\alpha)}$ part is central,
the generator $E$ commuting with the entire algebra).
  Introducing the Chern Simons invariant $S_{\hbox{cl}}^{(\alpha)}$ of
the flat connection we get
\eqn\oneloop{{\cal Z}({\cal
M})={(\hbox{det}(\Delta))^2\over|\hbox{det}(L_-)|}
\sum_{\alpha}e^{iS_{\hbox{cl}}^{(\alpha)}}
{\hbox{det}(L_-^{(\alpha)})\over(\hbox{det}(\Delta^{(\alpha)}))^2}}
The phase of  $\hbox{det}(L_-^{(\alpha)})$ is given, as in the $U(1)_N$ case,
 by
\eqn\phase{{i\over
4\pi}\int\epsilon_{\mu\nu\rho}A^{\mu}_E\partial^{\nu}A^{\rho}_E}
This term just combines with the classical Chern Simons action to incorporate
the appropriate quantum corrections \RsI\ , that amounts to replacing
\eqn\tilda{A_N\rightarrow A_{{\tilde N}}=A_N+{1\over 2k}A_E}
Up to the prefactor outside the sum,  the 3-manifold framing and spectral flow
contribution \Fg ,
  we thus get
 for each $\alpha$ the inverse of the analytic
 torsion
(instead of its square root in the $SU(2)$ case) as defined in \RayS, \WiI\ .
Owing to \RayS, \Mueller, \Cheeger\  this analytic torsion coincides
with the "algebraic" Reidemeister torsion
 as defined  in \ref\MilnorI{J.Milnor, Bull. Am. Math. Soc. 72 (1966) 358.}
and weighed by the exponential of
the Chern Simons action.
The inverse definition is also used  in the literature as in
 \Turaev\ and \Milnor and  this is what we shall do here.
For clarity we introduce the name \Turaev\ of Milnor torsion so
\oneloop\ becomes ${\cal
Z}\propto{1\over\tau_{\hbox{Mil}}(\hbox{trivial})}\sum\tau_{\hbox{Mil}}$
(and depending on the way this torsion is defined, we use the label
 analytic or
algebraic).
It is interesting to notice that
the partition function \oneloop\ depends on the global framing of the manifold
through the phase of  $\hbox{det}(L_-^{(0)})$. This dependence when interpreted
from a two dimensional point of view corresponds to a central charge $c=-2$.
On the other hand the $U(1,1)$ WZW model has $c=0$. The naive
correspondence between
Chern Simons and WZW theories, well known in the compact case,
does not hold here, illustrating \ref\Barnat{D.Bar Natan, E.Witten, Comm. Math.
Phys. 141 (1991) 423} (the $U(1,1)$ WZW model is still expected to
give rise to topological invariants
since such invariants can be shown to make
sense from a pure two dimensional conformal field theory point of view \Cr ,
\K , although the necessary axiomatics has presumably to be modified).
Fortunately the correct correspondence is easy to
establish, and does not spoil any of the results in \RsI , \RsII. One simply
has to change the coupling of one of the free bosons in the free field
representation of the $U(1,1)$ WZW model, and at the same time consider it as
antiholomorphic. The $S$ and $T$ matrices that follow are simply
deduced from the ones
in \RsII\  by $S\rightarrow -S$, $T\rightarrow\hbox{exp}(i\pi/3)T$, giving an
apparent central charge $c=-2$. All other results are unchanged.

Equation \oneloop\  is actually one loop exact.
This can be shown using a strategy
 similar to
 \ref\WiII{E.Witten, Nucl. Phys. B323 (1989) 113.}.
For simplicity forget for a while gauge fixing.  We compute
then  the partition function of some 3-manifold by first performing the
functional integration
over $A^{\mu}_N$, which produces a delta function
\eqn\ansum{\delta\left(\epsilon_{\mu\nu\rho}\partial^{\nu}A^{\rho}_E\right)}
so the integral over the variable $A_E$ is restricted to  $A_E$ flat
 connections
$F^{\mu\nu}_E=0$.
Assume as before that there is
a finite number of them $A_E^{(\alpha)}$. For each such connection
the integral over $A_{\psi}$ and $A_{\psi^{+}}$ is a  Gaussian
integral, so we get the ratio of the
determinant of the kinetic operator twisted by $A_E^{(\alpha)}$ over the
absolute value of the same
determinant but untwisted, where the denominator originates in the Jacobian
 for the constraint
$F^{\mu\nu}_E=0$.  Now let us take  gauge fixing into account.
 Infinitesimal  gauge transformations
can be written
\eqn\gauge{\eqalign{&\delta A_E^{\mu}=\partial^{\mu}\omega_E\cr
&\delta A_N^{\mu}=\partial^{\mu}\omega_N
+A^{\mu}_{\psi^{+}}\omega_{\psi}+
A^{\mu}_{\psi}\omega_{\psi^{+}}\cr
&\delta
A^{\mu}_{\psi}=\partial^{\mu}\omega_{\psi}+A_E^{\mu}\omega_{\psi}\cr
&\delta
A^{\mu}_{\psi^{+}}=\partial^{\mu}\omega_{\psi^{+}}-A_E^{\mu}\omega_{\psi^{+}}\cr
\cr}}
We choose the gauge fixing conditions
\eqn\gaugefixing{\partial^{\mu}A_E^{\mu}=\partial^{\mu}A_N
^{\mu}=D_{(E)}^{\mu}
A^{\mu}_{\psi}=D_{(E)}^{\mu}A^{\mu}_{\psi^{+}}=0}
where
\eqn\bigd{\eqalign{&D_{(E)}^{\mu}A^{\mu}_{\psi}=\partial^{\mu}
A^{\mu}_{\psi}+A^{\mu}_E
A^{\mu}_{\psi}\cr
&
D_{(E)}^{\mu}A^{\mu}_{\psi^{+}}=\partial^{\mu}A^{\mu}_{\psi^{+}}-A^{\mu}_E
A^{\mu}_{\psi^{+}}\cr}}
These conditions are implemented by adding a pair of bosonic and a pair of
fermionic Lagrange multipliers to the action:
$\phi^E,\phi^{N},\chi^{\psi},\chi^{\psi^{+}}$.
We also introduce  a pair of ghosts $b,c$ for each specie to restore
the correct
integration measure.
The ghost action contains
the following couplings
\eqn\ghostcoupl{\eqalign{&b^E\partial^{\mu}\partial^{\mu}c^E,\ 0,\ b^{\psi}(
\partial^{\mu}c^E)A^{\mu}_{\psi},\
 -b^{\psi^{+}}(\partial^{\mu}c^E)A^{\mu}_{\psi^{+}}\cr
&0,\ b^N\partial^{\mu} \partial^{\mu}c^N,\  0,\ 0\cr
&0\ ,b^N
\partial^{\mu} (A^{\mu}_{\psi^{+}}c^{\psi}),\ b^{\psi}D_{(E)}^{\mu}
D_{(E)}^{\mu}c^{\psi},\ 0\cr
&0,\ b^N
\partial^{\mu}(A^{\mu}_{\psi}c^{\psi^{+}}),\ 0,\ b^{\psi^{+}}D_{(E)}^{\mu}
D_{(E)}^{\mu}c^{\psi^{+}}\cr}}
Integration over $A_{N}$ produces again the condition \ansum\ . However
one has also to integrate over the Lagrange parameter
$\phi^E,\phi^{N}$ so the complete
Jacobian produces in denominator the absolute value
of the
determinant of untwisted $L_-$. Integration over $A_E$ is now restricted
to the sum over flat connections $A_E^{(\alpha)}$. We also
 integrate  over
$\chi^{\psi},\chi^{\psi^{+}}$ to ensure gauge conditions for
$A_{\psi},A_{\psi^+}$.  We can then integrate
 over the fermionic ghost $b^E$ to produce
a first determinant of
the
Laplacian $\Delta$ together with a $\delta$ function that restricts to  zero
modes of $c^E$ with respect to $\partial^{\mu}\partial^{\mu}$ .
 We have therefore  $\partial^{\mu}c^E=0$. Now
the integral over the bosonic ghost $b^{\psi}$ produces the
inverse of the determinant of the twisted Laplacian and restricts to
zero modes of $c^{\psi}$ with respect to $D^{\mu}_{(E)}D^{\mu}_{(E)}$, so we
have  $D_{(E)}^{\mu}c^{\psi}=0$. The term
$b^N
\partial^{\mu} (A^{\mu}_{\psi^{+}}c^{\psi})$ then vanishes identically
thanks to $D_{(E)}^{\mu}c^{\psi}=0$ and the gauge fixing condition on
$A_{\psi^{+}}^{\mu}$. Similar results are obtained for integration over
$b^{\psi^{+}}$. All non diagonal couplings thus disappear and we can finally
integrate over $b^N$. \ghostcoupl\ can therefore be reduced to its
diagonal, from which the correct  ratio of squares of
determinants of the Laplacian is immediately obtained. Having gotten rid of the
ghosts we
% undo the integration over $\chi^{\psi},\chi^{\psi^{+}}$
%and
integrate
over all remaining variables to produce determinant of twisted $L_-$ in
numerator.

Of course,
 integrations over $A_N^{\mu}$ and $A_E^{\mu}$
must be performed independently
for each
$U(1)_E$ and $U(1)_N$  bundles over the manifold.
Recall that $U(1)$ bundles are classified by maps from
 $H_1({\cal M})$ into $U(1)$. Also, when the bundle is non trivial,
 we must take into account the correct form
of the Chern Simons action in the manifold ${\cal M}$. Following
\ref\Djk{R.Dijkgraaf, E.Witten, Comm. Math. Phys. 129 (1990) 393.} it is easy
to show that, for a given bundle and its pair of $A_E,
A_N$ flat connections, the correct action for a gauge field $A$ is the
sum of the action of the flat connection $A^{(\alpha)}$ and the "naive" action
of
the difference $a=A-A^{(\alpha)}$. We thus recover the $S_{\hbox{cl}}$ phase
factors of \oneloop\ . We can then integrate freely over $a$ to recover
 the preceding
result in each bundle. Summing over all the bundles restores also
the  sum over $A_N$ flat
 connections that was missing so far,
so we  recover exactly \oneloop\ .

Our model behaves  much like the super $ISU(2)$ model in \WiII. In the latter
reference however the equivalent of
flat connections  $F^{\mu\nu}_E=0$ involves also a quadratic
term so the two determinants in \oneloop\ would both be twisted, and
  cancel each other,
giving
a  result simply equal to a phase  for each $\alpha$, that sums up to the
$SU(2)$ Casson invariant. Our model also bears some resemblance with $2+1$
dimensional gravity whose partition function would write \WiII ${\cal
Z}\propto\sum\left(\tau_{\hbox{Mil}}\right)^{-1}$.
 See also \ref\Bonac{G.Bonacina,
M.Martellini, M.Rasetti, "2+1 Dimensional Gravity as a Gaussian Fermionic
System and The 3D Ising Model", preprint POLFIS-TH-07-92.}.

The formula \oneloop\ requires corrections for possibly nontrivial
cohomologies $H^{0}_{\cal M}(A^{(\alpha)})$ and
$H^{1}_{\cal M}(A^{(\alpha)})$.
According to \Fg\ and \ref\WiIII{E.Witten, Comm. Math. Phys. 141 (1991) 153.}
a  nontrivial
$H^{0}_{\cal M}(A^{(\alpha)})$ requires an extra factor
of ${1\over\hbox{Vol}(H)}$, here $H$ is the subgroup of $U(1,1)$
commuting with all the holonomies of the connection $A^{(\alpha)}$.
The cohomology
$H^{0}_{\cal M}(A^{(\alpha)})$ is always nontrivial for a
$U(1,1)$ connection. Generally $H=U(1)_{E}\times U(1)_{N}$, for which (\RsII)
\eqn\volH{{1\over\hbox{Vol}(H)}=V={1\over 2k}}
However if the holonomy is a subgroup of $U(1)_{E}$ (i.e. if
$A^{(\alpha)}_{E}=0$ in a certain gauge, what we also call pure $A_N$ flat
connections), then $H=U(1,1)$ and
$\hbox{Vol}\left(U(1,1)\right)$ is equal to zero \RsII .
  We thus get a
diverging 3-manifold invariant that can nevertheless be compared with the
invariant of a reference manifold, say $S^3$. For $A_N$ flat connections,
the determinants (after proper subtraction of zero modes)
of twisted and untwisted operators coincide in \oneloop\ so
each of them contributes by a factor unity, as the Chern Simons action
vanishes. Other flat connections have a finite contribution, and thus are
"unobservable" in that case (but see later).
Therefore we simply count maps from $H_1({\cal M})$
into $U(1)_E$ with the result
\eqn\threemani{{{\cal Z}({\cal M})\over{\cal Z}(S^3)}=\hbox{order}(H_1({\cal
M}))}
where order=card here. This holds up to a phase factor encoding the global
framing of the manifold \RsII\ , and  as long as the $H_1$ of the manifold is
 finite. If it is infinite,
there is a continuous set of maps into $U(1)_E$ and the sum over flat
connections has to be replaced by an integral (see next paragraph).
 The final result is presumably
finite, so
\threemani\ should still hold if we define the order to be zero when
the cardinal is infinite:
\eqn\deforder{\hbox{order}=\hbox{card if card }<\infty,\ 0\hbox{ otherwise}}
The result \threemani\ then agrees with the numerical value of the
Milnor torsion
of ${\cal M}$ for the trivial representation of its $\Pi_1$ (see \Turaev).
Recall that in \RsII\ we recovered \threemani\ by surgery computations using
$S$ and $T$ matrices for the $U(1,1)$ WZW model. Taking into account the
modified correspondence between Chern Simons and WZW theories gives an
analogous result, up to a change of framing dependence $c=2\rightarrow c=-2$.

A nontrivial $H^{1}_{\cal M}(A^{(\alpha)})$
signals the existence of a continuous family of flat connections,
parametrized by a moduli space. This means that the sum in \oneloop\
has to be replaced  by an integral over that space.
$H^{1}_{\cal M}(A^{(\alpha)})$ takes values in the adjoint
representation of $U(1,1)$. Its bosonic part $\hbox{Bos}
H^{1}_{\cal M}(A^{(\alpha)}))$ is equal to the tensor square of the
ordinary cohomology of ${\cal M}$ over the real numbers:
\eqn\cohomology{\hbox{Bos}
(H^{1}_{\cal
M}(A^{(\alpha)}))=H^{1}_{\cal M}(R)\times H^{1}_{\cal M}(R)}
A particular example of a nontrivial $\hbox{Bos}(H^{1}_{\cal
M}(A^{(\alpha)}))$ is provided by the  manifold $S^{2}\times S^{1}$ and
will be considered later.

Let us finally discuss the fermionic flat connections. First one can easily
show that for any finite subgroup of $U(1,1)$, each element is conjugate of a
purely bosonic one. In the case of a Lens space
for instance, for which $\Pi_1=Z_p$ there are therefore only bosonic
 flat connections (up to gauge
equivalence) and the preceding discussion suffices. In general, if a
flat connection has  a fermionic component, the latter cannot be isolated,
which
implies a nontrivial fermionic part of $H^{1}_{\cal M}(A^{(\alpha)}))$
. In that case the fermionic integral cannot be saturated, and the contribution
to the partition function vanishes.
This agrees with the fact that nontrivial $\hbox{Ferm}(H^{1}_{\cal
M}(A^{(\alpha)})))$ corresponds to a zero mode in operator
$L_{-}^{(\alpha)}$ in \oneloop . We shall encounter such an example in the
study of Seifert manifolds.

\bigskip

Suppose now we consider a link $L$ in $S^3$. As discussed in \RsI\ and \RsII\
 using skein relations as well as surgery and $S$ and $T$ matrices, the
 $U(1,1)$ WZW
model should give rise to its multivariable Alexander polynomial. After taking
into account the proper Chern Simons WZW correspondence,
we expect a similar result to hold for
$U(1,1)$ Chern Simons theory. Let us
  comment on this, using now a pure  three dimensional point of view.

Recall first that  the definition of
the Alexander polynomial used here is the one adapted to the
multivariable case, ie we divide the generator of the Alexander
ideal by $t-1$. One has to set $t=\hbox{exp}(-2i\pi e/k)$ to match with
 quantum
field theory results \foot{The $t$ used here is  equal to the
square of the variable $t$ in \RsII }.
The knot complement has $H_1(S^3-K)=Z$ and the
variable $t$ of the Alexander polynomial corresponds to the generator of
this $H_1$ for algebraic  computation of torsion. Call $\Pi$ the infinite
cyclic multiplicative group with generator $t$. Recall
 the result of Milnor \Milnor\ that the Alexander polynomial
of a link in
$S^3$ is equal to the
"algebraic"  torsion
 of its complement (both objects being defined modulo $\Pi$ here).

As far as analytic torsion
is concerned let us go back to the Chern Simons point of view.
 We can cut out a tubular
neighborhood of the knot $K$ and compute the functional integral in
two steps.  First we integrate over the interior of the solid torus,
which results in a wave function for the gauge field  $A$ at its
boundary. Choose coordinates such that $dx^1$ is along the meridian,
$dx^2$ is along the longitude, $dx^0$ points inward. The holonomy of
$A^1$ being fixed \ref\Elitzur{S.Elitzur, G.Moore, A.Schwimmer,
N.Seiberg, Nucl.  Phys. B326 (1989) 108.} (and determining the flat
connection of the complement) we get the condition that the component
$a^1$ of the fluctuating part of
1-forms has to vanish at the boundary. To insure that $L_-$
is antihermitean and $\Delta$ is hermitean, we ask also that zero
 forms vanish at the boundary.  In the exterior the argument leading
to \oneloop\ and its exactness can be applied. We get therefore the
appropriate ratio of products of determinants
subject to these boundary conditions.
  Unfortunately they do not
reproduce the "absolute" or "relative" boundary conditions
  used in \RayS\ for which
the equality of analytic and algebraic torsion
was proven \Mueller\ , but rather are a
mixture of them. We suspect
however that the ratio of determinants in the analytic torsion is not
 too sensitive to  boundary
conditions.
 As an argument in that direction, notice,  as proven in \Milnor ,
that the chain complex $Q(t)\otimes_{\Pi} C_*({\tilde K},Z)$
(where $Q(t)$ is the field of rational functions in $t$ over
 rational numbers and
$\tilde{K}$
is the infinite cyclic covering of the complement), is acyclic. Accordingly
the algebraic torsion
\RayS\ is well defined modulo $\Pi$, independently of the metric chosen on
$S^3-K$ \RayS. Also the algebraic  torsion of the boundary is
 equal to one modulo $\Pi$ \Milnor\ . Therefore following \RayS\
 we expect that
the analytic torsions for relative or absolute boundary conditions should
 be equal up to powers of $t$.

It would be interesting to clarify the above discussion, in
particular concerning the role of $\Pi$.
If we assume  that boundary conditions
 inherited from Chern Simons
indeed  lead to the Milnor torsion, we see that the Milnor theorem is
equivalent to the exactness of the one loop computation in $U(1,1)$ Chern
Simons theory.

\bigskip

Now we would like to study further the torsion of manifolds.
Flat connections that are not pure $A_{N}$ give a finite contribution to the
partition function, but we do not know how to extract it in a
rigorous fashion from the above infinite background. There is however
a well defined procedure
that gives correct results: it consists in computing invariants of the manifold
with special links inside that acts as "observers" and factor out the pure
$A_N$ flat connections.
To explain this  let us first consider
the example of the lens
space $L(q,p)=X(p/q)$. Let us put a knot $K$ in it as follows: we start with
$S^{2}\times S^{1}$, remove a  solid torus $D\times S^{1}$
that contains a Wilson loop carrying the representation $(en/I)$,
and glue it back after
twisting the boundary by a $SL(2,Z)$ matrix with
 first column $p\choose q$. Suppose moreover $p/q$ has the simple continued
fraction expansion
\eqn\fraction{
p/q=a_2-{1\over a_1}}
and we suppose $a_1=q$ for simplicity (like in $L(q,-1)$,
otherwise take complex conjugates of the
final expressions).
To compute the corresponding invariant we follow the general strategy, ie apply
$T^{a_{2}}ST^{a_{1}}S$ to $|en/I>$ and evaluate scalar product of the final
state with $|n=0>$. Recall the $U(1,1)$ $S$ matrix elements
\eqn\smatrix{S_{en/I}^{e'n'/I}=-iV\hbox{exp}\left\{-{2i\pi\over
k}\left[e'(\tilde{n}-1/2)+e(\widetilde{n'}-1/2)\right]\right\}}
where $\tilde{n}=n+(e/2k)$, $V=1/(2k)$,
\eqn\smatrixI{S_{en/I}^{\widehat{n'}}=V{\hbox{exp}(-2i\pi en'/k)\over
2\hbox{sin}(\pi e/k)}}
Recall also the $T$ matrix
\eqn\tmatrix{T_{en/I}^{en/I}=\hbox{exp}2i\pi\left({(\tilde{n}-1/2)e\over k}
+{1\over 12}\right)}
One finds  \RsII
\eqn\zlink{\eqalign{&{\cal
Z}(L(q,p),K)=
-iV^{2}\hbox{exp}\left(i{\pi\over 6}(a_{1}+a_{2})\right)\cr
&\sum_{n'}\sum_{e'\neq
0}
{\hbox{exp}\left\{{2i\pi\over k}
[qe'(\widetilde{n'}-1/2)-e'(\widetilde{n}-1/2)-e(\widetilde{n'}-1/2)]
\right\}\over2\hbox{sin}(\pi e'/k)}\cr}}
The result depends therefore on the number of solutions in the fundamental
domain \RsII\ of the equation (for the unknown $e'$)
\eqn\modulo{
qe'=e \hbox{ mod }2k}
If there is no solution the invariant vanishes.

It is instructive to interpret the above result geometrically.
The empty Lens space has $\Pi_1=Z_q$ with a generator $g$ that satisfies
$g^q=1$. If we consider now the complement of the above knot $K$ in the Lens
space (obtained by removing a tubular neighborhood of $K$) it is easy to see
that it has $\Pi_1=Z$ with generator $t$ satisfying $g^q=t$. This complement is
in fact the second solid torus in the surgery prescription, ie it is the same
as for a simple loop $S^1$ embedded in $S^3$. However due to the twisting of
the boundary, "from the $S^3$ point of view", the generator of the $\Pi_1$ of
the complement is  $g$. Hence the invariant should be obtained by taking
the formula for the invariant in $S^3$ but replacing the $e$ charge by
$e'=e/q\hbox{ mod }2k$, as we indeed find. The phase term comes precisely from
the twisting of the boundary of the torus that turns into a non trivial self
framing here. In a "classical" theory of the Alexander polynomial, the
invariant of $K$ in the Lens space would not depend on arithmetic properties as
\modulo . In the quantum  theory however, charge quantization can make this
invariant vanish.

The above result is recovered using the classical computations of Alexander
invariants \Rolf . Represent the $\Pi_1$ of the knot complement as
\eqn\rep{\{t,g:\ tg^{-q}=1\}}
Then the matrix of Fox derivatives is \ref\FoxI{R.H.Cromwell, R.H.Fox,
"Introduction to knot theory", Springer Verlag (1963).}
\eqn\foxder{\left|\left|t^{-1},\ -t{t^{-1}-1\over t^{-1/q}-1}\right|\right|}
so the Alexander polynomial is, up to a power of $t$
\eqn\alex{\Delta_{\hbox{Alex}}\propto{1\over t^{1/q}-1}}

Notice that there is an ambiguity if we want to give a "numerical value" to
this
Alexander invariant, since several roots
 of $t$ can be chosen. This is
manifest in \zlink\ where the expression in the right hand side is periodic in
$e\hbox{ mod  }2k$, and can be interpreted again as a quantum effect.

Of course our purpose is not so much to compute invariants of links
 in manifolds but
to extract from them properties of the manifold itself, using the link as an
"observer", and letting ultimately the  $E$ charge carried by its Wilson loop
go to zero (mod $2k$) .  Notice that strictly speaking,
 this procedure is not allowed
in the quantum field theory where charges take discrete values and are limited
to the fundamental domain.  Let us clarify this.
Suppose there is a single
solution to \modulo\ : $e'=e/q$. Then the invariant is
\eqn\inva{
{\cal Z}(L(q,p),K)=-iV\hbox{exp}\left(i{\pi\over 6}(a_{1}+a_{2})\right)
{\hbox{exp}
\left[-{2i\pi\over k}{e\over q}(\tilde{n}-1/2)\right]\over 2\hbox{sin}(\pi
e/kq)}}

Let us take now the (formal)  limit $e=e_{0}\rightarrow 0$. One finds then
\eqn\limit{{\cal Z}(L(q,p))^{(0)}\rightarrow\
-\hbox{exp}\left(i{\pi\over 6}(a_{1}+a_{2})\right){iq\over 4\pi e_0}}
This agrees with the computation of the  invariant of an "empty" Lens
space \RsII\ : it is the order of the homology: $q$ ,
up to a framing factor and a
numerical factor that coincides with the invariant of $S^3$. \limit\ is of
course determined by pure $A_N$ flat connections only. Suppose now we take the
(formal) limit $e=e_0+2fk$ with $e_0\rightarrow 0$ and
 $f=0,\ldots,q-1$. In
that case one has
\eqn\minva{{\cal Z}(L(q,p),K)^{(f)}=-iV\hbox{exp}\left(i{\pi\over
6}(a_{1}+a_{2})\right)
\ {\hbox{exp}\left[
-{2i\pi\over k}{e_0+2fk\over q}(\tilde{n}-1/2)\right]
\over 2\hbox{sin}\left[\pi (e_0+2fk)/kq\right]}}
This has a finite limit for $f\neq 0$. Indeed the $E$ charge of the Wilson
loop constraints the holonomy $t$ and pure $A_N$ flat connections are projected
out. As we explain next the limit is then the Milnor torsion of the
appropriate flat connection.

To deal with a slightly more complicated case consider the Lens space
$L(p,-q)$ with $p/q$ same as  in \fraction .

We now put in $L$ two Wilson lines in the following fashion. We start with
$S^2\times S^1$, remove a solid tori $D\times S^1$ where we put a  loop
carrying some representation $|en/I>$, and glue it back after twisting  the
second $D\times S^1$ where we inserted a  loop carrying $|e'n'/I>$.
The partition
function is obtained by taking the scalar product of
$ST^{a_2}ST^{a_1}S|en/I>$ with $|e'n'/I>$. Only two
dimensional representations appear now in the intermediate states and the
computation looks much like in the $U(1)$ case.
Summing over intermediate states gives rise to arithmetic constraints as
before. Solve these constraints formally by taking their simplest solution as
above; this gives, up to framing of the manifold and factors depending on $k$
only
\eqn\res{\hbox{exp}\left\{{-2i\pi\over
k(a_1a_2-1)}\left[a_2e(\tilde{n}-1/2)+a_1e'(\widetilde{n'}-1/2)+
e'(\tilde{n}-1/2)+e(\widetilde{n'}-1/2)\right]\right\}}
Since $p=a_1a_2-1, q=a_1$ we can set  $a_2=q^*$ with
\eqn\qstar{q^*q=1\hbox{ mod }p}
{}From now on we set $e'=0$. As before we set
 $e=e_0+2fk,\ b=0,\ldots,p-1$
and we let $e=e_0\rightarrow 0$ go to zero to extract properties
 of the lens space itself
. We get
, collecting all factors,
\eqn\finitez{\eqalign{&{\cal Z}(L(p,q),K_1,K_2)^{(f)}=
iV\hbox{exp}\left({i\pi\over 6}(a_1+a_2)\right)\cr
&\hbox{exp}{-4i\pi q^*f^2
\over p}\hbox{exp}\left\{ {-2i\pi\over
p}\left[(2n-1)q^*f+(2n'-1)f\right]\right\}\cr}}
In the first exponential we recover the Chern Simons action of the $U(1)$ flat
connection associated with the representation of $\Pi_1=Z_p$ :
$g=\omega^f, \omega=\hbox{exp}(2i\pi /p)$.
 The $E$ charge being $0$ or a multiple of
$2k$ the Wilson loops behave as observers and simply project onto
a particular flat
connection of the Lens space; the second exponential then measures their
holonomies. For simplicity we have factored onto flat
connections with trivial $A_N$ part;  it suffices to consider $n=n_0+f'k$ to
extract the remaining ones, whose contributions differ from \finitez\ by phase
factors only (their torsion is the same).

Now \finitez\ is actually the partition function of
 the particular flat
connection times the product of two traces, one for each Wilson loop. We
therefore deduce the contribution of the flat connection itself
\eqn\torsion{iV\hbox{exp}\left({i\pi\over 6}(a_1+a_2)\right)
\ \hbox{exp}{-4i\pi q^*f^2
\over p}\ {1\over 4\hbox{ sin}{2\pi f\over p}\hbox{ sin}{2\pi q^*f\over p}}}
The factor $V$ occurs as explained earlier as the inverse of the volume of
$U(1)_E\times U(1)_N$. The first exponential occurs from
 the framing of the Lens space and the $i$ from spectral flow \Fg . The
second
 exponential gives  the Chern Simons action (there is no $k$ factor here
due to the form of \phase ). Forgetting for a while the torsion of the trivial
representation, and comparing with \oneloop\ we
conclude that  the fraction is equal to  the Milnor torsion of the  flat
connection
\eqn\lenstorsion{\tau^{(f)}={1\over |\omega^{2f}-1||\omega^{2q^*f}-1|}}
in agreement with the known results \ref\RaySII{D.B.Ray, I.M.Singer, in
Proceedings of AMS, 23 (1973) 167.}. A similar result follows from the above
calculation for $L(q,p)$ above. In that case, since only one loop has been
inserted in the Lens space
, one has to divide only by one
trace. However the other $\hbox{sin}$ term has already been
 provided by the $S$ matrix
 element, as
is usually the case for computations of Alexander invariants. Finally the
computation can be generalized to arbitrary continued fraction expansion as in
\RsII\ with similar results.

Dividing by the traces of the Wilson loops has also a torsion interpretation.
Indeed in our case we have the factorization $\tau(M)=\tau(M-K^b)\tau(K^b)$
 \Milnor ,
\ref\TuraevII{V.Turaev, Math. USSR Sbornik, 30 (1976) 221.}
 (where we used the fact
that the torsion of the boundary is one) and $K^b$ is a  tubular neighborhood
of $K$. $\tau(K^b)$ is the torsion of a solid torus, which is also the
complement of $S^1$ into $S^3$, and for which we know that the torsion (up to
phase factors) is equal to $\Delta_{\hbox{Alex}}\propto{1\over t-1}$.

 The complete  comparison with
\oneloop\
 involves an additional
factor equal to the torsion of the trivial flat connection, well defined only
after  a homology basis is chosen. With the usual choice, $\tau_0=p$
\Fg\
here.
We are missing this term in \torsion\ because
we did not treat the arithmetic constraints in the most correct fashion.

Let us now remedy this.
We first consider for a while the case of $U(1)$. For level $\kappa$
 we have the
$S$ and $T$ matrices
\eqn\Smat{S_{ab}={1\over\sqrt{\kappa}}
\hbox{exp}\left(-{2i\pi ab\over \kappa}\right),\
T_{aa}=\hbox{exp}i\pi\left({ a^2\over \kappa}-{1\over 12}\right)}
where $a,b=0,\ldots,\kappa-1$ and $\kappa$ is even.
 Notice that these matrix elements are invariant
under translations of $a,b\hbox{ mod }\kappa$. Let us now compute the invariant
of the Lens space $L(p,-1)$
 with two Wilson lines carrying representations $|a>$
and $|c>$ as above. We have
\eqn\zu{{\cal Z}(L(p,-1),K_1,K_2)=\hbox{exp}(-i\pi p/12)
{1\over\kappa}\sum_{b=0}^{\kappa-1}\hbox{exp}
\left[{i\pi\over\kappa}(-2ab+pb^2-2bc)\right]}
We would like to perform the resummation of \zu in a formal way,  without using
finite Gaussian sums formulas \Lisa\ . Due to the translation invariance
of the exponent in \zu\ we can extend the summation for $b$ over the entire
set $Z$. Let us also sum over $a\hbox{ mod }\kappa$. We write symbolically
(we suppress for a while the framing factor, which is all what is implied in
the symbol $\propto$)
\eqn\zuI{{\cal Z}(L(p,-1),K_1,K_2)\propto{1\over\kappa}{1\over\Lambda^2}
\sum_{n_a\in Z}\sum_{b\in Z}\hbox{exp}
\left[{i\pi\over\kappa}(-2ab+pb^2-2bc)\right]\hbox{exp}(-2i\pi n_ab)}
where $\Lambda$ are "normalization factors". The sum over $n_a$ however, due to
the last exponential, constrains by itself $b$ to be an integer. We can
therefore replace the second sum by an integral, dropping at the same time one
of the normalization factors
\eqn\zuII{{\cal Z}(L(p,-1),K_1,K_2)\propto{1\over\kappa}{1\over\Lambda}
\sum_{n_a\in Z}\int db\hbox{ exp}
\left[{i\pi\over\kappa}(-2(a+\kappa n_a)b+pb^2-2bc)\right]}
Compute now the Gaussian integral over $b$:
\eqn\zuIII{{\cal Z}(L(p,-1),K_1,K_2)\propto{i^{1/2}\over\sqrt{\kappa p}}
{1\over\Lambda}
\sum_{n_a\in Z}\hbox{exp}
\left[{-i\pi\over\kappa p}(a+c+\kappa n_a)^2\right]}
The exponential in \zuIII\ is invariant under shifts of $n_a\hbox{ mod }p$ so
we can split the sum over $n_a$ a sum over $f$ and a sum over $n$ with
$n_a=f+np$. The sum over $n$ produces a multiplicity cancelling the
normalization factor so we get
\eqn\zuIIII{{\cal Z}(L(p,-1),K_1,K_2)\propto{i^{1/2}\over\sqrt{\kappa p}}
\sum_{f=0}^{p-1}\hbox{exp}
\left[{-i\pi\over\kappa p}(a+c+\kappa f)^2\right]}
Let now $a=c=0$ to get (after reinstalling the framing factor)
\eqn\zuIIIII{{\cal Z}(L(p,-1))=\hbox{exp}(-i\pi p/12){i^{1/2}
\over\sqrt{\kappa p}}
\sum_{f=0}^{p-1}\hbox{exp}
\left({-i\pi\kappa f^2\over p}\right)}
This result can of course be recovered by a correct treatment of the
"normalization factors" ie by regularizing the sums.
The interpretation of \zuIIII\ is straightforward. We have again a
sum over $U(1)$ flat connections. The determinants are not twisted so as a
 global factor we get the inverse square root  of the
 torsion of the trivial flat connection. The factor of $\kappa$ as above comes
from the
volume of the group, since connections are reducible. We do not
have to divide by the contribution of the inserted loop
 in that case, since it is simply
equal to one, so we can suppress the label $K$ in the final result \zuIIIII .

This method  reproduces correct results in $SU(2)$ case
as we demonstrate now.
Consider again the
Lens space $L(p,-1)$ with a Wilson
 line carrying spin $j$ (in units where spins
are integer). We find
\eqn\zsu{\eqalign{&{\cal Z}(L(p,-1),K)=
\left({i\over 2p(k+2)}\right)^{1/2}\hbox{exp}\left({-i\pi p
\over 4}\right)\cr
&\sum_{f=0}^{p-1}e^{{-2i\pi(k+2)f^2\over p}}\left[
\hbox{cos}\left({2\pi fj\over p}\right)e^{{-i\pi j^2\over 2(k+2)p}}-
\hbox{cos}\left(
{2\pi f(j+2)\over p}\right)e^{{-i\pi (j+2)^2\over 2(k+2)p}}\right]\cr}}
If we set $j=0$ we recover the result of \Lisa\ (up to framing for which we
have put no corrections here)
. Let us instead consider the
limit $k\rightarrow\infty$. We get then
\eqn\zsuI{{\cal Z}(L(p,-1),K)\simeq\left({i\over
2pk}\right)^{1/2}\hbox{exp}(-i\pi p/4)
\sum_{f=1}^{[{p-1\over 2}]}e^{{-2i\pi kf^2\over p}}4\hbox{sin}
\left({2\pi f\over
p}\right)\hbox{sin}\left({2\pi f(j+1)\over p}\right)}
The first $\hbox{sin}$ in this formula is the inverse  of
the torsion of the complement. The second term can be explained as before. It
arises since the Wilson line creates some "small" (of order $1/k$)
background field through the holonomy condition. This field couples with the
"strong" (of order $k^0$) flat connection field proportional to $f$ so that
they contribute a bilinear term in the Chern Simons action. Due to the overall
normalization in this action, the final contribution does not depend on $k$.
The analog
 in the $U(1,1)$ case was a simple exponential: we get here a
 $\hbox{sin}$ due to
Weyl reflection, so in fact we have to deal with two holonomies of opposite
sign (see eg \Elitzur ). To get the contribution to the partition function of
the manifold itself we have to divide this result by the contribution of the
loop, which is the character of the spin
$j$ representation evaluated at the corresponding value of the holonomy
\eqn\charac{\hbox{ch}(\hbox{holonomy})={\hbox{sin}(2\pi f(j+1)/p)\over\hbox
{sin}(2\pi f/p)}}
We thus get in the end the contribution of the flat connection to be
$\hbox{sin}^2\left({2\pi f\over p}\right)$ as expected since the usual variable
$q^*=1$ here \foot{This is the
inverse of the Milnor torsion with our definitions.} . The sum in \zsuI
 runs only from
1 to integer part of ${p-1\over 2}$ due to "folding" of flat connections by
Weyl reflection. Notice the case $f=p$ does not contribute at this order. This
is because for $f=p$ one has to divide by the entire volume of the group, which
contributes additional factors of $k$ in the denominator. Appropriate factors
of $p$ appear also in that case.
 The torsion is also obtained from  \zsu\ or \zsuI\ by letting $j=0$. The point
is
we cannot put directly the identity representation in the Alexander case since
then one gets an expression dominated by pure $A_N$ flat connections.

We can now  get back to $U(1,1)$. The above resummation method applied formally
in that
case does not run into the arithmetic constraints of the above direct
 calculation; it can be considered as an ansatz to extract the finite
contributions to the partition function of manifolds. In the case of the Lens
space $L(p,q)$ we reproduce with this ansatz the result
\torsion\ together this time
with the contribution of the
trivial conection.  Let us  rather discuss  the more interesting
case of a Seifert manifold
$X(p_1/q_1,p_2/q_2,p_3/q_3)$. It is obtained from $S^2\times S^1$ by obvious
generalization of the construction of the Lens space $L(p,q)=X(q/p)$.
 Recall that
the $\Pi_1$ has a presentation $\{h,g_1,g_2,g_3;\ x^{p_i}h^{q_i}=1,\
x_1x_2x_3=1,\ h\hbox{ central}\}$, while
 $\hbox{order }(H_1(X))=p_1p_2q_3+\hbox{permutations}$ (we assume for
simplicity it is positive, otherwise one has to take the absolute value.)
 For further simplicity we restrict to $q_1=q_2=q_3=1$. Put in
each of the three torus $D^2\times S^1$ a Wilson loop carrying a representation
$(e_in_i/I)$. Put on the last torus $(en/I)$. We compute the invariant by
acting on each state with $T^{p_i}S$. As before constraints are met, which
solved formally would give (we suppressed  for a whileframing factors)
\eqn\seifert{\eqalign{&{\cal
Z}(X(p_1,p_2,p_3),L)\propto iV\hbox{exp}\left\{-{2i\pi\over k}
\left[{p_1p_2p_3\over (p_1p_2+p_1p_3+p_2p_3)}\left(\sum_i{e_i\over
p_i}+e\right)
   \right.\right.\cr
&\left.\left.\left({\sum_i\widetilde{n_i}-1/2\over p_i}+n-1/2\right)
-\sum_i{e_i(\widetilde{n_i}-1/2)\over p_i}\right]\right\}\cr
&\times\ \hbox{sin}^2\left[{2\pi\over k}
{p_1p_2p_3\over p_1p_2+p_1p_3+p_2p_3}\left(
\sum_i{e_i\over p_i}+e\right)\right]\cr}}
The detailed structure of $\Pi_1$ ie of
the flat conections depends now on the arithmetic
properties of $p_i$. Setting generally the $e_i$ to be appropriate multiples of
$2k$, we shall read the Chern Simons action and the holonomies in the above
exponential. The torsion will be obtained by dividing by a product of traces
for each Wilson loop, hence will have the form of
$\tau=\hbox{sin}^2/\hbox{sin}^4$, as is known. To get further results
 we can apply
the resummation ansatz.
After computations one finds for instance that, if $p_2+p_3$ and
$p_1p_2+p_1p_3+
   p_2p_3=P$ are coprimes,
and if we set $e_2=e_3=e=0$,
\eqn\zuu{\eqalign{&{\cal Z}(X(p_1,p_2,p_3),L)={iV\over
P}\hbox{exp}{i\pi\over 6}(p_1+p_2+p_3)\cr
&\sum_{f=0}^{P-1}\hbox{exp}\left({-4i\pi\over
P}(p_2+p_3)f^2\right)4\hbox{sin}^2\left({2\pi\over P}p_2p_3 f\right)\cr}}
which should completely describe the structure of flat connections in that
case, with torsion
\eqn\seitor{\tau^{(f)}=\left|
{\hbox{sin} (2\pi p_2p_3 f/P)\over 4 \hbox{ sin}(2\pi
(p_2+p_3)f/P)\hbox{ sin}(2\pi p_2 f/P)\hbox{ sin}(2\pi p_3 f/P)}\right|}
while the exponential gives the Chern Simons action and the prefactor the
torsion of the trivial representation equal to $\hbox{order }(H_1(X))=P$.
Notice that \seitor\ can sometimes vanish, indicating the existence of a zero
mode of $L_-^{(\alpha)}$. We have not found \seitor\ in the literature. It is
compatible with the torsions computed in \ref\Freed{D.Freed, J.Reine Angew.
Math. 429 (1992) 75.} and with various limiting cases.

\bigskip

Using $S$ and $T$ matrices, invariants of manifolds with an infinite $H_1$ have
also been computed in \RsII\ . The simplest case was
\eqn\sss{{\cal Z}(S^2\times S^1)=0}
(this is obtained before dividing by ${\cal Z}(S^3)$). The meaning of \sss\ in
terms of conformal blocks was disussed in \RsII\ . We can now
recover this result
from the torsion point of view by considering $S^2\times S^1$ as the
particular Lens space $L(0,1)$. Since $H_1(L)=Z$ there is a continuum of flat
connections parametrized by their holonomy along a noncontracible
cycle:
\eqn\h{h=\exp[i(En+Ne)]}
here $E$ and $N$ are the $U(1,1)$ generators defined in \rel , while
\eqn\oin{{2\pi e\over k}=\oint A^{\mu}_{E}dx_{\mu},\,\,
{2\pi n\over k}=\oint A^{\mu}_{N}dx_{\mu}}
The corresponding torsion is
\eqn\ta{\tau^{(en)}={1\over 4\sin^{2}{\pi e\over k}}}
Note that the manifold $S^{2}\times S^{1}$ can be formed by glueing
together two solid tori, so $\tau$ is the product of their
torsions.

The invariant of $S^{2}\times S^{1}$ is proportional to the integral of
the torsion over the moduli space of flat connections:
\eqn\zero{{\cal Z}(S^{2}\times S^{1})\propto\int^{2\pi}_{0}
dx{1\over \sin^{2}(x/2)}}
Such an integral appeared in \RsII\ in the  expression for the volume of
$U(1,1)$. We concluded there that this integral properly regularized has to be
set  equal to
zero, hence reproducing \sss\ .

The integrand in \zero\ is the  square of the "denominator" in the Weyl
character formula. It is also the  Jacobian factor for switching from the
integral over the group to the integral over its maximal torus. This
is not a coincidence: for a general bosonic Lie group $G$ the torsion for
$S^{2}\times S^{1}$ is
\eqn\Weil{\tau_{G}^{(\lambda)}=\prod_{\alpha>0}
|e^{i\alpha\cdot\lambda/2}-e^{-i\alpha\cdot\lambda/2}|^{2}}
where ${\lambda}=\oint A^{\mu}dx_{\mu}$  belongs to the
Cartan subalgebra. Indeed
 the  one loop approximation  of the
 Chern Simons path
integral over $S^{2}\times S^{1}$ with 2 Wilson lines carrying
representations $i$ and $j$ of $G$ along the noncontractible cycle,
expresses
orthonormality of Kac-Moody characters \WiI\ by
\eqn\char{\int_{m.t./W} d\lambda\ \tau_{G}^{(\lambda)}\chi_{i}(\lambda)
\chi_{j}(\lambda)=\delta_{i\bar{j}}}
where $m.t./W$ denotes a maximal torus factored over the action of the
Weyl group, and $\chi_{i,j}(\lambda)$ are the characters of the
representations $i$ and $j$.

More generally, it is interesting to consider the Milnor torsion
 of the manifold
$X_h\times S^1$ and compare it with the formula for the invariant obtained in
\RsII . By glueing and using the basic formula of the torsion of a solid torus,
one finds first of all  \WiIII , \Freed
\eqn\holes{\tau^{(en)}(X_h\times S^1)=\left|2\sin(\pi e/k)\right|^{2h-2}}
The one loop approximation to the path integral, which we expect to be exact
here, involves an integral over the moduli space of flat connections. We assume
that this can  be simply computed by integrating \holes\ over $e$, the
ratio of determinants in \oneloop\ becoming a measure on the moduli space \WiII
. One finds then
\eqn\hl{\int \tau^{(en)}=2^{2h-2}
\sum_{n=0}^{2h-2}{n\choose 2h-2}\int_0^{2\pi}e^{ix(h-1-n)} {dx\over 2\pi}=
2^{2h-2}{h-1\choose 2h-2}}
This binomial coefficient is easily interpreted in terms of conformal blocks.
It
is precisely the invariant obtained in \RsII , up to
some statistics factor. This additional factor arose in \RsII\ due to the
splitting in the quantization of the WZW model $|en>\rightarrow |en/1>,|en/2>$.
It is a puzzling problem  about the WZW CS correspondence
 to interpret such splitting from the three
dimensional point of view.

\bigskip

We finally would like to comment about $U(1,1)$ versus $GL(1,1)$. In many
respects, these two groups differ as their respective bosonic parts
 $U(1)\times U(1)$ and
$R\times R$ do. Their representations are similar; in the former case
 the numbers
$e,n$ are quantized, while they take continuous values in the latter. Both
groups have
a trivial $\Pi_3$ so the level $k$ is not quantized in the Wess Zumino
theories. Although the bosonic part of $U(1,1)$ is compact, in both cases the
corresponding  part in the Wess Zumino action has indefinite metric with
 signature
$(1,-1)$. Both theories lead to a spectrum unbounded from below. Most
computations in \RsI , in particular the computation of four point functions,
apply to both cases. $S$ and $T$ matrices have similar properties, although for
$GL(1,1)$ one has to integrate over $e,n$ in surgery computations (the
treatment of indecomposable and one dimensional representations is then quite
subtle). Strictly
speaking, the end of section 5 and section 7  in \RsI\ apply to $U(1,1)$ only
(this is unfortunately not indicated in the published version).
Considering now Chern Simons theories, both groups lead to the same invariant,
the Alexander polynomial, for links in $S^3$. Their difference appears
when dealing with invariants of links in more complicated manifolds. For
empty manifolds in particular, there is
$\left(\hbox{order}(H_1)\right)^2$
flat connections for $U(1,1)$, but only the  trivial
one  for $GL(1,1)$.
Up to the respective volumes of the groups, invariants  are equal
to  $\hbox{order}(H_1)$ in the former case, to $1$ in the latter.

\bigskip

Acknowledgments: we thank L.Crane, D.Auckley, D.Freed, G.Moore,
 V.Turaev  and E.Witten
for discussions. We especially thank D.Freed for kind explanations of his
work. HS was
supported by the Packard Foundation and by DOE contract DE-AC02-76ER03075. LR
was supported by the Robert A.Welch Foundation and by NSF Grant PHY9009850.

\listrefs\bye